\documentclass[10pt,oneside]{amsart}

\usepackage{tikz}
\usetikzlibrary{automata, positioning, arrows}

      \usepackage{amssymb}
\usepackage{amsfonts,amstext}
\usepackage{graphicx}
\usepackage{url}
\usepackage{amsmath,amstext,amssymb,amscd}
\usepackage{verbatim} 
\usepackage{mathtools}
      \usepackage{tabularx,booktabs}
      \usepackage{multicol}

\usepackage{amssymb,latexsym}
\usepackage{amsmath}
\usepackage{amsthm}
\usepackage{amssymb}
\usepackage{makecell}

\usepackage{adjustbox}
\theoremstyle{remark}

\theoremstyle{definition}

\usepackage{amsthm}
\allowdisplaybreaks

\usepackage{multicol}

      \makeatletter
      \def\@setcopyright{}
      \def\serieslogo@{}
     
      \makeatother

\begin{document}

\author{Adam Graham-Squire}
\address{Adam Graham-Squire, Department of Mathematical Sciences, High Point University, 1 University Parkway, High Point, NC, 27268}
\email{agrahams@highpoint.edu}

\author{David McCune}
\address{David McCune, Department of Physics and Mathematics, William Jewell College, 500 College Hill, Liberty, MO, 64068-1896}
\email{mccuned@william.jewell.edu}

\title[An Examination of Ranked Choice Voting in the United States, 2004-2022]{An Examination of Ranked Choice Voting in the United States, 2004-2022}

\begin{abstract}
 From the perspective of social choice theory, ranked-choice voting (RCV) is known to have many flaws. RCV can fail to elect a Condorcet winner and is susceptible to monotonicity paradoxes and the spoiler effect, for example. We use a database of 182 American ranked-choice elections for political office from the years 2004-2022 to investigate empirically how frequently RCV’s deficiencies manifest in practice. Our general finding is that RCV’s weaknesses are rarely observed in real-world elections, with the exception that ballot exhaustion frequently causes majoritarian failures.
\end{abstract}

 \subjclass[2010]{Primary 91B10; Secondary 91B14}

 \keywords{Condorcet winner, monotonicity paradox, spoiler effect, ballot exhaustion, majoritarian, empirical results}

\maketitle

\section{Introduction}

The use of ranked-choice voting (RCV) has greatly increased in the United States during the last few years. New York City first used RCV for city primary elections in 2021, the state of Maine has used it for state primary elections and elections for federal office since 2018, and the state of Alaska implemented RCV for federal and state offices in 2022. The voting method is well-known to have many deficiencies which receive attention in the social choice literature. The deficiencies with which we are concerned are:

\begin{itemize}
\item RCV can fail to elect the Condorcet winner.
\item  RCV is susceptible to the spoiler effect.
\item RCV is susceptible to downward and upward monotonicity paradoxes.
\item RCV is susceptible to the truncation paradox, the most extreme version of which is the no-show paradox.
\item RCV is susceptible to compromise strategic voting.
\item RCV is not truly “majoritarian” because of ballot exhaustion.
\end{itemize}

The purpose of this article is to examine how often these issues occur in actual elections, where we focus on the single-winner case. To that end, we collected the ballot data for as many single-winner ranked-choice American political elections as we could, resulting in a database of 182 elections. The flaws of RCV listed above can manifest only in elections without a majority candidate (i.e., elections in which RCV goes to at least a second round), and thus our database consists only of such elections. Our general finding is that these flaws occur rarely in actual ranked-choice elections, except for majoritarian failures caused by ballot exhaustion. While much of our analysis is new, perhaps the primary value of this article is that we provide a complete analysis of ``problematic'' American RCV elections, summarizing much of the empirical research of the last 18 years.

We note RCV has other potential flaws which are best examined through the lens of political science. For example, some of RCV’s detractors say that voters find the method confusing, or that RCV does not deliver on its proponents’ promise of creating more civil political campaigns. We do not address these claims; because of our backgrounds, we evaluate RCV only on criteria which have been examined in the mathematically-oriented social choice literature.

\section{Definitions: Ranked-Choice Voting and Its Deficiencies}

We begin with a definition of RCV and then formally define the flaws with which we are concerned, with examples throughout.  Note that the social choice literature generally uses the term \emph{RCV} to refer to any election that involves ranking candidates, and the literature uses terms such as instant runoff voting, alternative vote, the Hare rule, etc., to describe the vote method defined below. We use the term RCV, as it aligns with how the term used by most municipal and state elections offices in the US.

In an RCV election, voters cast preference ballots which allow a voter to rank the candidates in order of preference. Voters often do not provide a complete ranking, either by choice or because the voters’ jurisdiction limits the number of candidates that can be ranked on a ballot. For example, the city of Minneapolis, MN allows voters to rank only three candidates, regardless of how many candidates are in the race. After an election the ballots are aggregated into a \emph{preference profile}, which shows the number of each type of ballot cast. For example, Table \ref{AK_election} shows a preference profile for the August 2022 Special Election for the single US House seat in Alaska, an election involving the three candidates Nick Begich, Sarah Palin, and Mary Peltola\footnote{The race also included several write-in candidates, all of whom received a relatively trivial amount of votes. We eliminated these candidates before creating the preference profile.}. The table shows that 27070 voters ranked Begich first, Palin second, and Peltola third; the other numbers across the top row convey similar information about the number of voters who cast the corresponding type of ballot\footnote{There is some ambiguity about how the ballots were processed, and our preference profile numbers differ by a handful of ballots from the official vote counts. For example, according to Alaska Division of Elections the 27070 should be 27053. These small differences do not affect our conclusions.}. We use the notation $A > B$ to denote that a voter ranks candidate $A$ above $B$, and thus 27070 voters choose the ranking Begich $>$ Palin $>$ Peltola. To keep the table a manageable size, we combine ballots of the form $A > B$ and $A > B > C$, which has no effect on the winner of the election under RCV. We also combine ballots of these forms in all subsequent examples. Note that voters are not required to provide a complete ranking of the candidates, and some voters choose to rank only a single candidate on their ballots. For example, 11262 voters ranked Begich on their ballots but did not rank Palin or Peltola.
\begin{table}[]
  \centering

  \begin{adjustbox}{width=\textwidth}

\begin{tabular}{l|ccccccccc}
Num. Voters & 27070   &15478&11262 & 34078& 3659 &21237&47407&4647&23733\\
\hline
1st choice & Begich & Begich & Begich & Palin & Palin &Palin&Peltola &Peltola &Peltola\\
2nd choice & Palin & Peltola & $-$ & Begich &Peltola &$-$&Begich&Palin&$-$\\
3rd choice & Peltola & Palin & $-$ & Peltola & Begich &$-$&Palin&Begich&$-$\\
\end{tabular}

 \end{adjustbox}
  \caption{The August 2022 Alaska Special Election for US House, after eliminating write-in candidates.}
  \label{AK_election}
\end{table}

The method of RCV declares a winner as follows: if the election contains a \emph{majority candidate}, a candidate who receives a majority of first-place votes, then that candidate is declared the winner. Otherwise eliminate the candidate(s) with the fewest first-place votes and transfer that candidate’s votes to other candidates based on the second-place choices on that eliminated candidate’s ballots. If a surviving candidate now has a majority of the remaining votes, that candidate wins; otherwise, the process of elimination and vote-transfer continues in this manner until a candidate has secured a majority of (first-place and transferred) votes.

We illustrate the RCV algorithm using the preference profile in Table \ref{AK_election}. Begich receives 53810 first-place votes, Palin receives 58974, and Peltola receives 75799, and thus Begich is eliminated. As a result, 27070 of his votes are transferred to Palin, 15478 are transferred to Peltola, and 11262 are dropped from the election (these ballots are said to be \emph{exhausted}). After the vote transfer, Peltola wins the election with 91277 votes to Palin’s 86044.

If an election contains more than three candidates then RCV proceeds similarly, but the process may take more rounds to select a winner. We now define the flaws of RCV that we investigate in this article.

\textbf{Condorcet Failure}: In the Alaska House election, if we compare Begich to Palin in a head-to-head matchup then 101229 voters prefer Begich to Palin whereas 63621 voters prefer Palin to Begich. Similarly, 93052 voters prefer Begich to Peltola whereas 79558 voters prefer Peltola to Begich. Based on the ranking information provided by the voters\footnote{We choose to interpret ballots using the weak order model (Popov et al., 2014), in which all candidates left off a ballot are tied for the lowest ranking on a voter’s ballot.}, Begich wins each of his head-to-head matchups and thus is the Condorcet winner of the election. Condorcet winners receive much attention in the social choice literature because they are considered ``consensus'' or ``strong'' candidates, candidates who ``should'' win (assuming the election contains such a candidate). We say that an RCV election demonstrates a \emph{Condorcet failure} when the election contains a Condorcet winner and RCV fails to select this candidate, which occurs in the Alaska House election.

\textbf{Spoiler Effect}: An RCV election demonstrates the \emph{spoiler effect} if there exists a subset of losing candidates such that removing these candidates from the election causes the winner of the election to change. In the Alaska House election, if the losing candidate Palin were removed then Begich would win, and thus this election demonstrates a spoiler effect under RCV.

\textbf{Upward Monotonicity Paradox}: An RCV election demonstrates an \emph{upward monotonicity paradox} if there exists a set of ballots such that shifting the RCV winner up the rankings on those ballots but keeping the relative rankings of the other candidates the same, thereby creating a hypothetical second preference profile in which the winner has more voter support, creates an election in which the original RCV winner does not win. In the Alaska House election, if 6000 ballots on which Palin is ranked first and no other candidate is ranked on the ballot were changed to ballots of the form Peltola $>$ Palin then Peltola would lose the resulting election. That is, giving the winning candidate Peltola more first place votes by shifting her up the rankings on some ballots turns her into a loser. The reason is that even though Peltola picks up an additional 6000 first-place votes in the hypothetical second election, this additional support causes Palin to be eliminated from the election first (in contrast to the original election, in which Begich was eliminated first) and then Peltola would lose to Begich in the final round.

\textbf{Downward Monotonicity Paradox}: An RCV election demonstrates a \emph{downward monotonicity paradox} if there exists a set of ballots and a losing candidate $L$ such that shifting $L$ down the rankings on those ballots but keeping the relative rankings of the other candidates the same, thereby creating a hypothetical second preference profile in which this losing candidate has less voter support, creates an election in which $L$ is the RCV winner. Our running example with the Alaska data does not demonstrate a downward paradox: if either Begich or Palin were shifted down on any set of ballots, they would remain losers of the election.

To demonstrate a real-world example of this paradox, consider the 2020 Board of Supervisors election for the 7th District in San Francisco, CA. The election contained seven (not write-in) candidates; Table \ref{SF_election} shows the preference profile for the election after four candidates have been eliminated and their votes transferred. At this point in the RCV process, Engardio has 14119 first-place votes, Melgar has 11652, and Nguyen has 10855. Nguyen is eliminated, and after their votes are transferred Melgar wins with 18561 votes to Engardio’s 16370. If Engardio were shifted down one ranking on 800 ballots of the form Engardio $>$ Nguyen $>$ Melgar so that those ballots change to Nguyen $>$ Engardio $>$ Melgar then in the resulting election Melgar would be eliminated first and, even with less voter support, Engardio would still have enough votes to defeat Nguyen head-to-head in the final round. That is, shifting the losing candidate Engardio down on some voters’ ballots would turn Engardio into the RCV winner. As with the Alaska election, this example creates a paradoxical outcome by changing the order in which candidates are eliminated.

\begin{table}[]
  \centering

  \begin{adjustbox}{width=\textwidth}

\begin{tabular}{l|ccccccccc}
Num.Voters & 5237   &3316&5566 & 4050& 5708 &1894&2251&6909&1695\\
\hline
1st choice & Engardio & Engardio & Engardio & Melgar & Melgar &Melgar&Nguyen &Nguyen &Nguyen\\
2nd choice & Melgar & Nguyen & $-$ & Engardio&Nguyen &$-$&Engardio&Melgar&$-$\\
3rd choice & Nguyen& Melgar & $-$ &Nguyen & Engardio&$-$&Melgar&Engardio&$-$\\
\end{tabular}

 \end{adjustbox}
  \caption{The 2020 District 7 Board of Supervisors election in San Francisco, CA, after eliminating all but the final three candidates.}
  \label{SF_election}
\end{table}

\textbf{Truncation Paradox}: An election demonstrates a \emph{truncation paradox} if there exists a set of voters such that the voters could create a more desirable electoral outcome by ranking fewer candidates on their ballots; i.e., when these voters’ express less information on their ballots then a candidate whom they prefer more wins the election than if the voters expressed more information. The most extreme version of this paradox is a \textbf{no-show paradox}, which occurs when there exists a set of voters such that removing their ballots from the election creates a more desirable electoral outcome for those voters (removing the ballots altogether is the most extreme form of ballot truncation). To be precise, a no-show paradox occurs if $A$ is the RCV winner and there exists a losing candidate $B$ and a set of voters that prefer $B$ to $A$ such that if these voters abstain from voting then $B$ would be the winner in the modified election with these voters’ ballots removed. The Alaska election in Table \ref{AK_election} demonstrates this paradox: if 5400 voters who cast the ballot Palin $>$ Begich $>$ Peltola were removed from the election then Palin would be eliminated first (recall that in the original election, Begich was eliminated first) and, despite the removal of these ballots, Begich would have enough support to defeat Peltola head-to-head in the final round. These 5400 voters would have achieved a more desirable outcome (their second-favorite candidate would have won instead of their third favorite) if they had abstained.

\textbf{Compromise Voting Failure}: \emph{Compromise voting} is a form of strategic voting in which some voters calculate that their favorite candidate cannot win and so insincerely rank another candidate as their first choice. This plays out regularly in plurality elections, where voters often cast a vote for a candidate perceived as viable instead of voting for their favorite. Generalizing the definition from (Green-Armytage, 2014), we say that an election demonstrates a \emph{compromise voting failure} if there exists a losing candidate $A$ and a set of ballots such that $A$ is ranked above the original RCV winner, and $A$ becomes the RCV winner if we shift them up to the first ranking on these ballots. That is, this failure occurs if there exists a set of voters who should have cast a “compromise vote” for $A$, thereby causing $A$ to win. The definition in (Green-Armytage, 2014) is much more restrictive: they consider only elections in which shifting $A$ to the top of \emph{all} ballots on which $A$ is ranked over the original RCV winner turns $A$ into the winner.

The Alaska election in Table \ref{AK_election} demonstrates this failure: if 5400 voters who ranked Palin first and Begich second had ranked Begich first instead then Begich would have won the election and these voters would have obtained their second choice instead of their third. To obtain a more desirable electoral outcome, these voters should have ``compromised'' for Begich. This election also fits the more restrictive definition from (Green-Armytage, 2014), as Begich would become the winner if all 34078 voters who cast the ballot Palin $>$ Begich $>$ Peltola were to rank Begich first.

\textbf{Majoritarian Failure}: We say that an election demonstrates a \emph{majoritarian failure} if, when the RCV algorithm is run until there are only two candidates left, the winning candidate in this final round does not achieve a majority of the total number of votes cast. Both the Alaska and San Francisco elections demonstrate this failure. Table \ref{AK_election} shows a total of 188583 voters in that election (which increases slightly if we were to include write-in candidates), yet the winner Peltola earns 91277 in the final round, achieving only 48.4\% of the votes of the total electorate. Similarly, in the San Francisco election there were 39322 ballots cast and the winner Melgar earns $18561/39322 = 47.2\%$ of the total vote in the final round. The reason these winners fail to earn a majority is \emph{ballot exhaustion} (Burnett and Kogan, 2015), where many partial ballots are discarded before the final round because these ballots do not rank either of the two final candidates. We use the term \emph{winner’s vote share} to refer to the percentage of the total votes earned by the RCV winner in the final round when there are only two candidates remaining, so that Melgar’s vote share is 47.2\%, for example.

\section{Data Sources and Collection}

We collected the vote data for as many American single-winner ranked-choice political elections as we could, with the restriction that we obtained data only for elections in which no candidate earns an initial majority of first-place votes. In total our database contains 182 elections, 147 of which were collected by the first author for (McCune \& McCune, 2022a). Most of the elections are for municipal office such as mayor or city councilor; a handful of elections are for statewide or federal elections, such as elections for US House and Senate in Maine. The data was collected from election office websites, and some was received by request from election offices when the data was not posted. After our initial round of data collection, the RCV advocacy organization FairVote created a publicly accessible repository of American ranked-choice data (Otis, 2022) and we collected a handful of additional elections from this source. Table \ref{data_sources} gives a summary of our database, including the number of elections from each jurisdiction and the years for which data is available.
\begin{table}[]
  \centering

\begin{tabular}{c|c|c}
\textbf{Jurisdiction}&\textbf{Years for which we have data}&\textbf{Num. Elections}\\
\hline

Alaska & 2022 & 13\\
Aspen, CO & 2009 & 2\\
Berkeley, CA & 2010, 2014-2020 & 6\\
Bloomington, MN & 2021 & 1 \\
Burlington, VT & 2006, 2009 & 2\\
Corvallis, OR & 2022 & 2\\
Easthampton, MA & 2020 & 1\\
Eastpointe, MI & 2020 & 1\\
Elk Ridge, UT & 2021 & 1 \\
Las Cruces, NM & 2019 & 3\\
Maine & 2018-2022 & 11\\
Minneapolis, MN&2009, 2013, 2017, 2021 & 26\\
Minnetonka, MN&2021&1\\
New York City, NY & 2021 & 41\\
Oakland, CA & 2010-2020 & 21\\
Pierce County, WA & 2008-2009 & 4\\
Portland, ME & 2021-2022 & 2\\
San Francisco, CA & 2004-2022 & 32\\
San Leandro, CA & 2010-2014 & 5\\
Santa Fe, NM & 2018 & 2\\
Springville, UT & 2021 & 1\\
St. Louis Park, MN & 2019, 2021 & 2\\
Telluride, CO & 2015 & 1\\
Woodland Hills, UT &2021 & 1\\

\end{tabular}

  \caption{Summary of data sources.}
  \label{data_sources}
\end{table}

\section{Prior Social Choice Literature on RCV’s Flaws}

There is a vast social choice literature which evaluates RCV as a voting method, and most of this literature is theoretical. Due to the empirical nature of our work we do not attempt to survey the theoretical literature; instead, we focus on prior empirical work.

\emph{Condorcet failures}: The largest empirical investigation of Condorcet failures in American political elections occurs in (McCune \& McCune, 2022a), which analyzed 147 of the elections in our database. The authors found only one Condorcet failure, in a previously documented election from Burlington, VT (Gierzynski et al. 2010; Ornstein \& Norman 2014). (McCune \& McCune, 2022a) also found only one single-winner election without a Condorcet winner. (Graham-Squire \& Zayatz, 2021) analyze 35 elections from our database, again finding a failure in only the Burlington election, and (Song, 2022) analyzed many\footnote{We are not sure how many of the elections analysed in (Song, 2022) contain majority candidates and so we cannot determine how much of their work overlaps with ours.} of the pre-2021 elections in our database for Condorcet failures. RCV advocacy groups such as FairVote also check ranked-choice data for Condorcet failures (Landsman 2017; Otis 2021). 

There is a small literature about Condorcet failures which investigates elections outside the context of American political elections. For example, (Regenwetter et al. 2007; Popov et al. 2014) analyse ranked-choice election data from the American Psychological Association, using the data to generate tens of thousands of election pseudoprofiles. They find an extremely low rate of Condorcet failures under RCV. (Darmann et al. 2019) use survey data collected prior to the 2015 parliamentary elections in the Austrian federal state of Styria, and again find little evidence of Condorcet failures in the data.

\emph{Spoiler Effect}: Most of the discussion of spoilers has been limited to the theoretical literature, with the exception that many non-academic articles discuss potential spoilers in individual elections. For example, many newspaper articles (for a typical example example, see (Bokat-Lindell, 2021)) mention the famous case of the 2000 US Presidential election in which Ralph Nader is commonly understood to have spoiled the election for Al Gore. (McCune \& Wilson, 2022) is the largest empirical study of the spoiler effect in American ranked-choice political elections. The authors find two ranked-choice elections demonstrating the spoiler effect out of 170 analyzed, 147 of which are in our database. They also perform bootstrap analysis to generate pseudoprofiles, similar to the analysis in (Popov et al. 2014), and find low rates of the spoiler effect in the generated data. A Condorcet failure is a special case of the spoiler effect as we have defined it, and thus the Condorcet studies mentioned above also indirectly address the spoiler effect.

\emph{Monotonicity and truncation paradoxes}: Most of the elections in our database have not been previously processed by code which searches for monotonicity or truncation paradoxes. The largest empirical study of monotonicity and truncation paradoxes in American political ranked-choice elections occurs in (Graham-Squire \& Zayatz, 2021), which analyses elections from Alameda County and San Francisco, CA, as well as a mayoral election from Burlington, VT. The authors find a single upward monotonicity paradox and no truncation paradoxes out of 35 elections without a majority candidate. (McCune \& Graham-Squire, 2022) analyze 1079 ranked-choice multiwinner elections from Scotland and find low (but non-zero) rates of upward monotonicity, downward monotonicity, and no-show paradoxes. Many articles which analyze real-world monotonicity failures focus on single elections. See (Gierzynski et al. 2010; Ornstein \& Norman 2014) for an analysis of the 2009 Burlington, VT mayoral election and (McCune \& McCune, 2022b) for an analysis of monotonicity failures in a 2021 city council election in Minneapolis, MN. 

	Some prior work is semi-empirical in that the authors use polling or survey data to estimate monotonicity failure rates in real-world elections. For examples of this kind of analysis see (Gallagher, 2013), which studies single-transferable vote elections in Ireland, and (Miller, 2017), which provides a semi-empirical analysis of English general elections from 1992-2010.
	
\emph{Compromise voting failures}: We are unaware of empirical studies which study this failure. (Green-Armytage, 2014) and (Green-Armytage et al. 2016) analyze compromise voting using a variety of models of voter behavior, and generally find that RCV is much less susceptible to this issue than other famous voting methods such as plurality and the Borda count.

\emph{Majoritarian failures}: The only empirical studies of majoritarian failures of which we are aware are (Burnett and Kogan, 2015), which analyze four American elections in our database, and (Kilgour et al. 2020), which analyze 18 ranked-choice elections, 4 of which are in our database (the other 14 contain majority candidates or are elections from the UK or the American Psychological Association). FairVote also seems to have done substantial analysis of ballot exhaustion\footnote{See \url{https://fairvote.org/resources/data-on-rcv/#evaluating-rcv-election-outcomesnbsp}.}, but it is unclear which elections were analyzed. 

More generally, many studies address the topic of partial ballots without focusing on majoritarian failures per se. (Coll, 2021) and (Donovan et al. 2022) study which demographic groups are more likely to cast partial ballots. (Tomlinson et al., 2022) analyse the effects of ballot truncation on the number of possible winners in ranked-choice elections. 

\section{Methodology for Detecting RCV Weaknesses in Each Election}

All elections in the database were processed using Python code which searched for the given flaw. The code is adapted from programs used in (Graham-Squire \& Zayatz 2021), (McCune \& McCune 2022a), (McCune \& Wilson 2022), and (McCune \& Graham-Squire 2022). Checking if the RCV winner is the Condorcet winner or if the RCV winner earns a majority of the initial total votes in the final round is computationally straightforward. However, there are challenges when searching for the spoiler effect, monotonicity paradoxes, truncation paradoxes, or compromise voting failures.
	
	If the number of candidates in an election is large enough then, due to limits of computation time, we cannot check every subset of losing candidates for a change in the winner when this subset is dropped. For example, the 2013 Minneapolis mayoral election contained 35 candidates, resulting in billions of possible sets of candidates to check for the spoiler effect. For all elections in the database we checked for individual spoiler candidates, but for elections with large numbers of candidates we additionally checked only candidate subsets of size two. For elections with more than twelve candidates, we also ran the RCV algorithm until only ten candidates remained and then checked the resulting election for the spoiler effect. Most of the elections were already processed in this manner in (McCune \& Wilson, 2022).
	
	To demonstrate a monotonicity or truncation paradox, we must find a set of ballots such that shifting a candidate up or down the rankings, or truncating the ballots, causes a change in the order of elimination which causes a paradoxical change in the winner. Except for the case of three-candidate elections in which every voter provides a complete ranking, there are no known necessary and sufficient conditions for an election to exhibit one of these paradoxes. Thus, if an election exhibits such a paradox, we cannot guarantee our code will find it. However, our monotonicity and truncation code has been thoroughly means-tested in other projects and has successfully found many elections which demonstrate monotonicity paradoxes. For example, the code found 21 elections demonstrating upward monotonicity paradoxes in Scottish local government elections which had not been previously documented (McCune \& Graham-Squire, 2022). The code essentially works by strategically changing or truncating ballots to achieve a change in the order of elimination or the candidates, and measures if a paradoxical outcome occurs as a result of the ballot changes.
	
	The code we use to search for compromise voting failures is a straightforward adaptation of our monotonicity code.

\section{Results}

We now present our results, separating the issue of ballot exhaustion from the others. Only eight elections in the database, listed below, exhibit any kind of non-majoritarian flaw. Four of the elections demonstrate only a compromise voting failure; we include the preference profile for only one of these elections as the dynamics of the other three are similar.

\textbf{2009 Mayoral Election in Burlington, VT (Table \ref{Burlington_election}).} This election demonstrates the following flaws (Gierzynski et al. 2010; Ornstein \& Norman 2014):

\begin{itemize}
\item Condorcet failure: the Condorcet winner, Montroll, is not the RCV winner, Kiss.
\item Spoiler effect: If the losing candidate Wright were removed from the election, the winner changes from Kiss to Montroll.
\item Upward monotonicity paradox: If 450 voters who voted only for Wright and 300 voters who ranked Wright first and Kiss second shift Kiss up to the first ranking on their ballots, then Kiss would no longer win the election.
\item Compromise voting failure: If all voters who ranked Montroll over Kiss but did not rank Montroll first were to ``compromise'' and rank Montroll first then the RCV winner would be Montroll instead of Kiss. (This failure has not been pointed out previously.)
\end{itemize}

\begin{table}[]
  \centering

  \begin{adjustbox}{width=\textwidth}

\begin{tabular}{l|ccccccccc}
Num.Voters & 2043   &371&568 & 1332& 767 &455&495&1513&1289\\
\hline
1st choice & Kiss & Kiss & Kiss &Montroll & Montroll &Montroll&Wright &Wright&Wright\\
2nd choice & Montroll & Wright & $-$ & Kiss &Wright &$-$&Kiss&Montroll&$-$\\
3rd choice & Wright & Montroll & $-$ & Wright & Kiss &$-$&Montroll&Kiss&$-$\\
\end{tabular}

 \end{adjustbox}
  \caption{The 2009 mayoral election in Burlington, VT, after eliminating all but the final three candidates. This table is taken from (Ornstein \& Norman, 2014).}
  \label{Burlington_election}
\end{table}

\textbf{2009 County Executive Election in Pierce County, WA (Table \ref{Pierce_election})}. This election demonstrates a compromise voting failure. In the actual election McCarthy defeated Bunney 136346 votes to 132292 in the final round. If 15000 voters who did not rank Goings first but ranked Goings above McCarthy were to rank Goings first then Goings would be the RCV winner. To see why, note that in Table 5 the first-place vote totals for Bunney, Goings, and McCarthy are 118690, 77417, and 92208, respectively. The vote gap between Bunney and McCarthy, the two candidates who advance to the final round, is $118690-92208=26482$, while the gap between McCarthy and Goings is $92208-77417=14791$. Thus, in this round the RCV winner McCarthy is closer to being eliminated than to having the most votes, and if we can find a number of voters between 14792 and 26481 who rank Bunney first and rank Goings over McCarthy (and the data does contain these voters) then we can shift Goings up to first on these ballots so that McCarthy is eliminated. If the gap between McCarthy and Bunney were smaller than the gap between McCarthy and Goings then we could not make this failure occur, as McCarthy would advance to the final round no matter how we construct compromise votes. Thus, the compromise voting failure in this election is “non-monotonic” in some sense because Goings benefits from the extra support of 15000 voters but does not benefit from the extra support of more than 26482.

\begin{table}[]
  \centering

  \begin{adjustbox}{width=\textwidth}

\begin{tabular}{l|ccccccccc}
Num.Voters & 27661   &27375&63654 & 13602& 44138 &19687&12330&59502&20376\\
\hline
1st choice & Bunney & Bunney & Bunney &Goings &Goings &Goings&McCarthy &McCarthy&McCarthy\\
2nd choice & Goings& McCarthy& $-$ & Bunney &McCarthy &$-$&Bunney&Goings&$-$\\
3rd choice & McCarthy& Goings & $-$ & McCarthy& Bunney &$-$&Goings&Bunney&$-$\\
\end{tabular}

 \end{adjustbox}
  \caption{The 2008 County Executive election in Pierce County, WA, after eliminating all but the final three candidates.}
  \label{Pierce_election}
\end{table}

\textbf{2010 Mayoral Election in Oakland, CA.} This election demonstrates a compromise voting failure. In the actual election Quan was the RCV winner. If 2400 voters who did not rank Kaplan first but ranked Kaplan over Quan were to rank Kaplan first then she would be the RCV winner.

\textbf{2016 District 2 City Council Election in Berkeley, CA.} This election demonstrates a compromise voting failure. In the actual election Davila was the RCV winner. If 130 voters who did not rank Armstrong-Temple first but ranked Armstrong-Temple over Davila were to rank Armstrong-Temple first then she would be the RCV winner.

\textbf{2017 Ward 3 City Council Election in Minneapolis.} This election demonstrates a compromise voting failure. In the actual election Fletcher was the RCV winner. If 370 voters who did not rank Bildsoe first but ranked Bildsoe over Fletcher were to rank Bildsoe first then he would be the RCV winner.

\textbf{2020 District 7 Board of Supervisors Election in San Francisco, CA (Table \ref{SF_election})}. This election demonstrates a downward monotonicity paradox (as shown above), which has not been previously documented.

\textbf{2021 Ward 2 City Council Election in Minneapolis, MN (Table \ref{Minneapolis_election})}. This election demonstrates the following flaws (McCune \& McCune, 2022b).

\begin{itemize}
\item Spoiler effect: Worlobah is the RCV winner of the election. If the losing candidate Arab were removed from the election, the winner changes to Gordon.
\item Upward monotonicity paradox: If 456 of the voters who ranked Arab first and Worlobah second shift Worlobah up one ranking, Worlobah would lose the election.
\item Downward monotonicity paradox: If 80 of the voters who ranked Arab first and Gordon second shift Arab down one ranking, Arab would win the election.
\item Compromise voting failure: If the voters who did not rank Gordon first but did rank Gordon above Worlobah were to rank Gordon first then Gordon would be the RCV winner. (This failure has not been pointed out previously.)
\end{itemize}

This election cannot demonstrate a Condorcet failure as we have defined it because there is no Condorcet winner. All other elections in the database contain a Condorcet winner.

\begin{table}[]
  \centering

  \begin{adjustbox}{width=\textwidth}

\begin{tabular}{l|ccccccccc}
Num.Voters &  801  &1177&822 & 908& 756 &1572&1299&1088&492\\
\hline
1st choice & Gordon & Gordon & Gordon & Arab & Arab &Arab&Worlobah &Worlobah &Worlobah\\
2nd choice & Arab& Worlobah & $-$ & Gordon &Worlobah &$-$&Gordon&Arab&$-$\\
3rd choice & Worlobah & Arab & $-$ & Worlobah & Gordon &$-$&Arab&Gordon&$-$\\
\end{tabular}

 \end{adjustbox}
  \caption{The 2021 Ward 2 city council election in Minneapolis, MN, after eliminating all but the final three candidates.}
  \label{Minneapolis_election}
\end{table}

\textbf{August 2022 Alaska Special Election for US House (Table \ref{AK_election})}. This election demonstrates the following flaws, as demonstrated above (Graham-Squire \& McCune, 2022).

\begin{itemize}
\item Condorcet failure.
\item Spoiler effect.
\item Upward monotonicity paradox.
\item No-show paradox.
\item Compromise voting failure. (This failure has not been pointed out previously.)
\end{itemize}

Of the six elections which demonstrate a compromise voting failure, only three (Alaska, Burlington, and the 2021 Minneapolis election) demonstrate this failure in the strong sense of the definition from (Green-Armytage, 2014).

Of the 182 elections in our database, 95 demonstrate a majoritarian failure. That is, 95 elections have the property that when we run the RCV algorithm until only two candidates remain, the winning candidate does not secure a majority of the total votes cast in the first round.

\section{Discussion}

As the results of the previous section suggest, anomalies other than majoritarian failures occur very rarely in real-world ranked-choice elections (see Table \ref{failure_rates}). Recall that none of our elections contain majority candidates; we did not attempt to count the number of American political ranked-choice elections which contain such a candidate, but there are easily at least 100 such elections across all jurisdictions. If we were to include these elections, the failure rates become significantly smaller. Thus, non-majoritarian failures seem to be of little practical concern in real-world elections. We note, however, that even one failure could potentially have large consequences. The city of Burlington, VT repealed the use of RCV after the 2009 mayoral election, for example. Also, if a failure were observed in an election for a very important office then RCV’s overall good performance becomes less important. For example, the state of Maine uses RCV to allocate its Electoral College votes in US Presidential elections; a Condorcet failure or monotonicity paradox in such an election would likely have much more weight than when these failures occurred in a city council election in Minneapolis.

Monotonicity paradoxes, truncation paradoxes, and compromise voting failures are all specific cases of RCV being susceptible to strategic voting; i.e., all of these failures show that RCV is manipulable. Our results show that such manipulability is rarely a concern in practice; furthermore, we argue the data shows that even in elections which are manipulable in some way, it would be very difficult for voters to implement tactical voting successfully. It is hard to believe that voters would vote insincerely to attempt to engineer a monotonicity or no-show paradox; these paradoxes occur rarely enough and affect a relatively small enough number of voters that attempting to manipulate an election in this fashion seems like an absurd strategy. Compromise voting failures occur more frequently, but the RCV algorithm is complicated enough that it is not clear that groups of voters could correctly anticipate when to cast a compromise vote. For example, in the 2009 County Executive Election in Pierce County, WA (Table \ref{Pierce_election}), voters who ranked Bunney first and ranked Goings over McCarthy would have to anticipate that Bunney could not defeat McCarthy head-to-head in the final round and would have to calculate that in the penultimate round the gap between Goings and McCarthy would be smaller than the gap between McCarthy and Bunney. Furthermore, if more than 26482 of these voters decide to make this compromise then McCarthy would still win because this level of compromising would cause Bunney to be eliminated and McCarthy would advance to the final round and defeat Goings. Thus, there is a relatively narrow range of vote compromising that would allow Goings to win the election. It does not seem likely that voters would make these calculations. We also note that our definition of a compromise voting failure is quite expansive, and thus it is notable that the failure rate is so low.

\begin{table}[]
  \centering

\begin{tabular}{c|ccccccc}
\textbf{Flaw}&Condorcet & Spoiler & Upward & Downward &  Truncation & Compromise&Majoritarian\\
\hline
\textbf{Rate}&1.1\%&1.6\%&1.6\%&1.1\%&0.5\%&3.8\%&52.5\%\\
\end{tabular}

  \caption{The failure rate in our database of 182 elections for the six non-majoritarian flaws of RCV. For the Condorcet failure rate we use a denominator of 181 because one of the elections does not contain a Condorcet winner.}
  \label{failure_rates}
\end{table}

The only election in which we think voters might have been able to calculate correctly that they should cast compromise votes is the August 2022 Alaska House election, which had unique political dynamics. The election contained only three candidates, two Republicans (Begich and Palin) and one Democrat (Peltola), and Palin had a national profile which made her a polarizing figure. Peltola was a moderate, non-polarizing candidate, and thus voters who cast the ballot Palin $>$ Begich $>$ Peltola could probably anticipate that Palin would not defeat Peltola head-to-head. Therefore, such voters should have been able to calculate that their only chance of electing a Republican to the House was to cast a compromise vote for Begich, but these voters seemingly did not cast compromise votes. Furthermore, this House election was an off-schedule special election which occurred because of the death of the sitting congressman and this House seat was up for election again in November 2022, just three months later. The November election contained four candidates, the same three from the August election and a fourth candidate Chris Bye, but Bye received less than 2\% of the vote and thus the November election was essentially just a rerun of the August election. In response to the issue-riddled August election, supporters of Begich and Palin did not seem to meaningfully alter their behaviour. Palin still received approximately 5000 more votes than Begich, but because Peltola increased her support by a substantial relative amount (most likely due to higher turnout in this general election), she was the Condorcet winner and defeated Palin by a wider margin in the final round. Furthermore, the November election contained none of the RCV failures we discuss. Thus, even when voters are given poll data of the highest quality, an election which occurred three months prior, they do not seem to react in a strategic fashion (at least, the number of voters reacting strategically seems relatively small).

In summary of non-majoritarian failures, while RCV is manipulable in theory it does not seem to be manipulable in practice. Non-majoritarian failures occur infrequently and, when they do occur, it is not clear that these failures are ``actionable'' on the part of voters. Issues such as monotonicity paradoxes are undesirable, but they become offensive only in hindsight; we find little evidence that voters could vote strategically to engineer such outcomes in actual elections. Our empirical results are consonant with the theoretical work of (Green-Armytage 2014) and (Green-Armytage et al. 2016), which show that RCV is less manipulable than most other voting methods.

Majoritarian failures occur at a much higher rate than the other failures, accounting for more than half the elections in the database. These occur because a significant portion of the electorate in these elections cast partial ballots, causing their ballots to become exhausted before the final round. As pointed out in (Burnett and Kogan, 2015), partial ballots occur for two reasons. First, voters may voluntarily provide an incomplete ranking, choosing not to rank all candidates. Second, some jurisdictions limit the number of candidates that voters can rank on their ballots. For example, the city of Minneapolis, MN, allows voters to rank only three candidates regardless of the number of candidates in the race. In such elections, voters are often forced to cast partial ballots.

For our discussion, it is useful to distinguish between elections in which voters can provide a complete ranking of the candidates if they so choose, and elections in which they cannot. In an election in which the jurisdiction limits the number of candidates ranked on ballots, we say that the election’s \emph{truncation level} is the number of candidates a voter can rank. If the number of candidates in an election is more than the election’s truncation level plus one, we say the election is \emph{truncated} because voters cannot provide a complete ranking of the candidates. For example, a ranked-choice election in Minneapolis with five or more candidates is truncated; if an election contains four or fewer candidates, voters can provide a complete ranking\footnote{In an election with four candidates and truncation level 3, voters can provide a complete ranking of the four candidates by providing a complete ranking of their top three; the candidate left off the ballot is assumed to be the voter’s fourth choice.}. In our database 72 elections are truncated and 110 are not.

Because voters are forced to cast partial ballots in truncated elections and a majoritarian failure is caused by a significant portion of ballots being partial, we expect these failures to be more common in truncated than non-truncated elections. Our findings bear this out: 57 of the 72 truncated elections demonstrate a majoritarian failure, while 38 of the 110 non-truncated elections demonstrate this flaw. In elections with a majoritarian failure, the winner’s vote share in truncated elections also tends to be lower (on average, 43.9\%) than the winner’s vote share in the non-truncated elections (on average, 47.7\%). Table \ref{trunc} shows the five elections with the smallest winner vote share among the truncated elections. These vote shares are all significantly lower than 40\%, quite far from a majority. The 2010 San Francisco Board of Supervisors election in the 10th district is particularly extreme, with the winner earning less than 25\% of the total vote. By contrast, Table \ref{non-trunc} shows the five elections with the smallest winner’s vote share among the non-truncated elections. The smallest winner’s vote share among the non-truncated elections is 42.2\%; there are 13 truncated elections in which the winner’s vote share is smaller.
\begin{table}[]
  \centering

\begin{tabular}{c|c}
\textbf{Election}&\textbf{Winner's Vote Share}\\
\hline
2010 San Francisco Board of Supervisors Dist. 10 & 24.3\%\\
2021 NYC Dem. Primary City Council Dist. 9& 35.2\%\\
2020 Minneapolis City Council Ward 6 & 36.1\%\\
2021 NYC Dem. Primary Borough President Kings & 37.3\%\\
2008 Pierce County, WA, County Treasurer & 37.5\%
 \end{tabular}

  \caption{The five truncated elections with the smallest winner vote shares.}
  \label{trunc}
\end{table}

\begin{table}[]
  \centering

\begin{tabular}{c|c}
\textbf{Election}&\textbf{Winner's Vote Share}\\
\hline
2021 NYC Rep. Primary Dist. 50 & 42.2\%\\
2021 Portland, ME City Council At-Large& 44.5\%\\
2019 San Francisco District Attorney & 44.9\%\\
2021 NYC Dem. Primary Dist. 32 & 45.6\%\\
2008 Pierce County, WA, County Executive & 45.6\%
 \end{tabular}

  \caption{The five non-truncated elections with the smallest winner vote shares.}
  \label{non-trunc}
\end{table}

The high rate of majoritarian failure in the data seems concerning. There are a few potential solutions to address this failure rate. First, jurisdictions with truncated elections could remove the limit on the number of candidates, allowing voters to provide a complete ranking. Our results suggest this would have an effect on the failure rate, and we are unaware of mathematical reasons for including a truncation level in an election (although there may be political reasons for doing so). Second, some have argued (Kilgour et al. 2020) that voters may cast partial ballots due to the cognitive load of trying to rank multiple candidates. If this is the case, jurisdictions could try to alleviate this load by finding ways to limit the number of candidates on the ballot. For example, ranked-choice elections in Alaska contain at most four candidates because prior to the RCV election there is a primary election which uses plurality voting to whittle down the field of candidates. However, in the 2022 ranked-choice elections in Alaska we still see a high rate of majoritarian failure, with six out of thirteen elections demonstrating this issue (although the smallest of the winner vote shares is 47.2\%, and so these failures are not particularly egregious). It is possible that voters will be willing to rank more candidates over time as they become more comfortable with RCV, in which case Alaska’s strategy of limiting the number of candidates to four could significantly lower the rate of majoritarian failures.

Of course, depending on one’s values it is possible that majoritarian failures are not important, and therefore the high failure rates are irrelevant. In non-truncated elections, if voters who cast partial ballots are voting sincerely then it is possible that there does not exist a candidate who could earn majority support in a final round of RCV. For example, if enough voters care only about one or two candidates and are indifferent among the rest, then no voting method will be able to find a ``majority winner,'' and it is not the fault of RCV that the election produces a majoritarian failure. However, even if one finds majoritarian failures unimportant, given the high rate of such failures (even in non-truncated elections) it is likely advisable that RCV advocates adjust their rhetoric around RCV. For example, (Lair 2022) states that ``the majoritarian principle is an axiom of democratic government'' and uses this statement (among others) to justify the adoption of RCV. Similarly, (Lavin 2019) states: ``$[$W$]$e need majority rule in elections—not only as a principle or best practice but as a practical assurance to legitimize outcomes and give elected officials strong mandates to govern. By requiring winners to earn a majority of votes—if not in first choices alone then with backup choices—RCV meets both of these critical needs.'' Our results suggest that RCV does not live up to such statements in practice.

\section{Conclusion}

When evaluated based on criteria important to the social choice literature, RCV mostly performs well in practice. In the American ranked-choice political elections in our database, RCV almost always selects the Condorcet winner and avoids the spoiler effect, while also demonstrating practical resistance to strategic voting. Paradoxes which feature prominently in the theoretical literature such as monotonicity and no-show paradoxes seem to occur on the order of 0.5-1.1\% for real-world elections without a majority candidate, and these failure rates would decrease considerably if we also included ranked-choice elections which do not advance to a second round. The percentage of elections in which the winner does not receive a majority in the final round is very high, which should give pause to RCV advocates.

	Since a perfect voting method seemingly does not exist, choosing a method involves trade-offs. The weaknesses of RCV are mostly not observed in real-world ranked-choice data available in the US. Of course, the failure rates are not zero, and it is reasonable to insist on a method which always chooses the Condorcet winner or is not susceptible to monotonicity paradoxes. We have contributed to the literature by providing a comprehensive empirical analysis of the social-choice weaknesses of RCV in the US, but whether RCV’s benefits outweigh its costs is, in our view, still an open question.

\section*{Acknowledgements}

Thank you to Deb Otis for showing us the FairVote data repository.

\begin{thebibliography}{99}


\bibitem{BL} Bokat-Lindell, S. (2021). “Can Ranked-Choice Voting Cure American Politics?,” 
New York Times, June 24, 2021, \url{https://www.nytimes.com/2021/06/24/opinion/ranked-choice-new-york.html. Accessed Jan. 14, 2023.}
\bibitem{BK} Burnett C. \& Kogan V. (2015). Ballot (and voter) ``exhaustion'' under Instant Runoff Voting: An examination of four ranked-choice elections. \emph{Electoral Studies}, 37, 41-49.


\bibitem{C} Coll J. (2021). Demographic Disparities Using Ranked-Choice Voting? Ranking Difficulty, Under-Voting, and the 2020 Democratic Primary. \emph{Politics and Governance}, 9 (2), 293-305.
\bibitem{DGK} Darmann A., Grundner J., \& Klamler C. (2019).  Evaluative voting or classical voting rules: Does it make a difference? Empirical evidence for consensus among voting rules. \emph{European Journal of Political Economy}, 59, 345-353.
\bibitem{DTH} Donovan T., Tolbert C., \& Harper S. (2022). Demographic differences in understanding and utilization of ranked choice voting. \emph{Social Science Quarterly}, 103 (7), 1539-1550.
\bibitem{G} Gallagher M. (2013). Monotonicity and non-monotonicity at PR-STV elections. Paper presented at the annual conference of the elections, public opinion and parties (EPOP) specialist group.
\bibitem{GHS} Gierzynski A., Hamilton W., \& Smith W. Burlington Vermont 2009 IRV mayor election. \url{https://rangevoting.org/Burlington.html}. Accessed 12/20/22.
\bibitem{GSM} Graham-Squire A. \& McCune D. (2022). A Mathematical Analysis of the 2022 Alaska Special Election for US House. Preprint: \url{https://arxiv.org/abs/2209.04764}.


\bibitem{GSZ} Graham-Squire A. \& Zayatz N. (2021). Lack of Monotonicity Anomalies in Empirical Data of Instant-runoff Elections. \emph{Representation}, 57 (4), 565-573.
\bibitem{GA} Green-Armytage J. (2014). Strategic voting and nomination. \emph{Social Choice and Welfare}, 42, 111-138.

\bibitem{GATC} Green-Armytage J., Tideman T.N., \& Cosman R. (2016). Statistical evaluation of voting rules. \emph{Social Choice and Welfare}, 46, 183-212.
\bibitem{KGF} Kilgour D.M., Grégoire J.C., \& A.M. Foley. (2020). The prevalence and consequences of ballot truncation in ranked-choice elections. \emph{Public Choice} 184: 197-218.


\bibitem{Lair} Lair S. (2022). Ranked-choice voting is needed, along with open primaries. \url{https://thenevadaindependent.com/article/ranked-choice-voting-is-needed-along-with-open-primaries/}. Accessed 12/26/22.
\bibitem{Landsman} Landsman T. (2017). All RCV Elections in the Bay Area So Far Have Produced Condorcet Winners. \url{https://fairvote.org/every_rcv_election_in_the_bay_area_so_far_has_produced_condorcet_winners/}. Accessed 12/26/22.
\bibitem{Lavin} Lavin N. (2019). Majority Rule: more than just a principle for successful elections. \url{https://fairvote.org/majority_rule_more_than_just_a_principle_for_successful_elections/}. Accessed 12/26/22.
\bibitem {MGS} McCune D. and Graham-Squire A. (2022). Monotonicity Anomalies in Scottish Local Government Elections. Preprint.
\bibitem{MMa} McCune D. \& McCune L. (2022). Does the Choice of Preferential Voting Method Matter? An Empirical Study Using Ranked Choice Elections in the United States. \emph{Representation}. \url{https://doi.org/10.1080/00344893.2022.2133003}
\bibitem{MMb} McCune D. \& McCune L. (2022). The Curious Case of the 2021 Minneapolis Ward 2 City Council Election, to appear in \emph{The College Mathematics Journal}.


\bibitem{Mi} Miller N.R. (2017). Closeness matters: Monotonicity failure in IRV elections with three candidates. \emph{Public Choice} 173 (1-2): 91-108.
\bibitem{O1} Otis D. RCV in New York City. (2021, October 14).  Retrieved from \url{https://www.fairvote.org/rcv_in_new_york_city#candidate_analysis}.
\bibitem{O2} Otis D. (2022). Single winner ranked choice voting CVRs. \url{https://doi.org10.7910/DVN/AMK8PJ}, Harvard Dataverse, V5.


\bibitem{ON} Ornstein J. \& Norman R. (2014).  Frequency of monotonicity failure under Instant Runoff Voting: estimates based on a spatial model of elections. \emph{Public Choice}, 161 (1-2), 1-9.
\bibitem{PPR} Popov S., Popova A., \& Regenwetter M. (2014).  Consensus in Organizations: Hunting for the Social Choice Conundrum in APA Elections. \emph{Decision} 1 (2), 123-146.
\bibitem{RKH} Regenwetter M., Kim A., \& Ho M. (2007). The Unexpected Empirical Consensus Among Consensus Methods. Psychological Science 18 (7), 629-635.
\bibitem{S}Song C.G. (2022). Three Empirical Analyses of Voting $[$Unpublished doctoral dissertation$]$. Virginia Tech.
\bibitem{TUK}Tomlinson K., Ugander J., \& Kleinberg J. Ballot Length in Instant Runoff Voting. Preprint:\url{https://arxiv.org/abs/2207.08958}.


\end{thebibliography}
\end{document}